\documentclass[utf8]{arxiv_Harvard} 
\usepackage{url,hyperref,lineno,microtype,subcaption}
\usepackage[onehalfspacing]{setspace}
\usepackage{float}
\usepackage{threeparttable}
\usepackage{multirow}
\usepackage{makecell}
\usepackage{colortbl}
\usepackage{graphicx}

\def\keyFont{\fontsize{8}{11}\helveticabold }
\def\firstAuthorLast{Guo {et~al.}} 
\def\Authors{Liyuan~Guo\,$^{1,*, \dagger}$, Annika~Weiße\,$^{1,\dagger}$, Seyed~Mohammad~Ali~Zeinolabedin\,$^{2}$, Franz~Marcus~Sch\"uffny\,$^{1}$, Marco~Stolba\,$^{1}$, Qier~Ma\,$^{1}$, Zhuo~Wang\,$^{1}$, Stefan~Scholze\,$^{1}$, Andreas~Dixius\,$^{1}$, Marc~Berthel\,$^{1}$, Johannes~Partzsch\,$^{1}$, Dennis~Walter\,$^{1}$, Georg~Ellguth\,$^{1}$, Sebastian~H\"oppner\,$^{1}$, Richard~George\,$^{1}$ and Christian~Mayr\,$^{1}$}
\def\Address{
$^{1}$
Faculty of Electrical and Computer Engineering, School of Engineering Sciences, Dresden University of Technology, Germany \\
$^{2}$
Department of Electrical and Computer Engineering, College of Engineering, University of Utah, Salt Lake City, UT, United States
}
\def\corrAuthor{Liyuan Guo}

\def\corrEmail{liyuan.guo@tu-dresden.de}

\begin{document}
\onecolumn
\firstpage{1}

\title[Implantable Neural Signal Processing PSoC]{68-Channel Highly-Integrated Neural Signal Processing PSoC with On-Chip Feature Extraction, Compression, and Hardware Accelerators for Neuroprosthetics in 22nm FDSOI} 

\author[\firstAuthorLast ]{\Authors}
\address{}
\correspondance{}

\extraAuth{$^{\dagger}$These authors share first authorship.}

\maketitle

\begin{abstract}

\section{}
Multi-channel electrophysiology systems for recording of neuronal activity face significant data throughput limitations, hampering real-time, data-informed experiments. These limitations impact both experimental neurobiology research and next-generation neuroprosthetics. We present a novel solution that leverages the high integration density of 22nm FDSOI CMOS technology to address these challenges.
The proposed highly integrated programmable System-on-Chip comprises 68-channel 0.41~\textmu W/Ch recording frontends, spike detectors, 16-channel 0.87-4.39~\textmu W/Ch action potential and 8-channel 0.32~\textmu W/Ch local field potential codecs, as well as a MAC-assisted power-efficient processor operating at 25~MHz (5.19~\textmu W/MHz). The system supports on-chip training processes for compression, training and inference for neural spike sorting. The spike sorting achieves an average accuracy of 91.48\% or 94.12\% depending on the utilized features.
The proposed PSoC is optimized for reduced area (9~mm$^2$) and power. On-chip processing and compression capabilities free up the data bottlenecks in data transmission (up to 91\% space saving ratio), and moreover enable a fully autonomous yet flexible processor-driven operation. Combined, these design considerations overcome data-bottlenecks by allowing on-chip feature extraction and subsequent compression. 

\tiny
 \keyFont{ \section{Keywords:} biomedical signal processing, neural recording system, digital integrated circuits, neural signal compression, implantable devices, biomedical electronics, spike sorting} 
\end{abstract}

\section{Introduction}\label{sec:intro}
Recent advances in the analysis of neuronal microcircuits and the design of neural prosthetic devices, demand the analysis of neural activities recorded with high spatial and temporal resolution from hundreds or even thousands of channels \citep{GEORGE2020101589}.
However, transmitting such a substantial volume of neural signals off-chip proves to be highly power-consuming, constrained by bandwidth limitations, and poses challenges in terms of storage capacity, rendering some significant applications of extracellular signals such as brain-machine interfaces and neural prosthetics challenging. 

Conventional neural signal acquisition SoCs typically consist of analog frontends (AFEs) responsible for amplifying, filtering, and digitizing raw data. Here, an increase in electrodes causes a proportional rise in throughput, memory, and power requirements. Analysis from \cite{zeinolabedin2022} suggests that a conventional 1000-channel recording system requires a power budget of approximately 250 mW for off-chip transmission of raw neural signals. Yet, to ensure thermal biocompatibility of an implantable system, power consumption must remain below 35 mW, as demonstrated by the 3D Utah electrode array \citep{kim2008}. Furthermore, bandwidth analysis reveals the limitations of conventional acquisition approaches. For instance, assuming a sampling frequency of 20 kHz and each sample comprising 9 bits, a recording system with 1000 channels necessitates a bandwidth of at least 180 Mbps. Consequently, integrating on-chip digital processing engines becomes imperative.

\begin{figure}[t]
\centerline{\includegraphics[width=85mm,keepaspectratio]{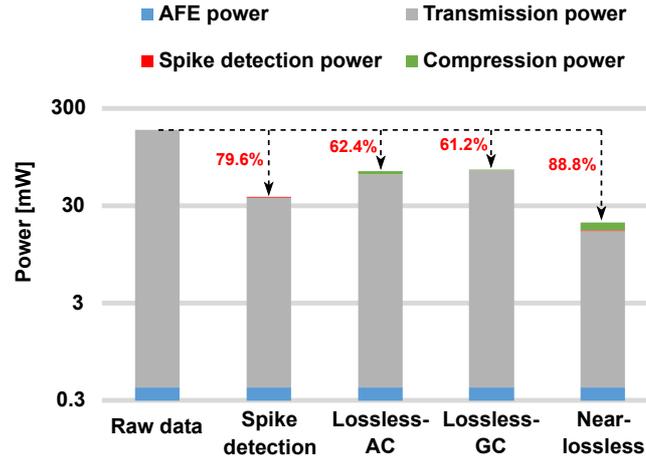}}
\caption{Power analysis of a 1000-channel neural recording system. For the proposed PSoC: P(AFE)=0.41 \textmu W/Ch, P(Lossless-AC)=4.39 \textmu W/Ch, P(Lossless-GC)=0.87 \textmu W/Ch, P(Near-lossless)=3.24 \textmu W/Ch, firing rate=60 spikes/sec, transmission energy=1 nJ/bit, sampling frequency=20 kHz, and the data bit-width=9 bits. AC represents arithmetic coding, while GC denotes Golomb coding.}
\label{fig:pwr_analysis}
\end{figure}

With a focus on alleviating these limitations, a crucial subset of on-chip digital processing engines is dedicated to spike processing, specifically spike detection and spike sorting. If the average neuronal firing rate is approximately 60~-~100 spikes per second, with each spike having a window size of 64 samples under a sampling rate of 20~kHz, spike detection would theoretically reduce the data rate to about 36.2~-~59.6 Mbps for a 1000-channel recording system. This includes the transmission of inter-spike intervals and channel indexes to ensure data reconstruction. This reduction in data rate translates to a potential power reduction of about 66.6\%~-~79.6\% compared to conventional recording systems, as shown in Figure~\ref{fig:pwr_analysis}. Spike sorting has the potential to further reduce data rate and power consumption by transmitting only the spike index instead of the entire spike. Reference \cite{chae2009} reports a spike detection and feature extraction for spike sorting, albeit limited to processing a single channel at a time. In \cite{karkare2013}, a 16-channel SoC performing training and inference on-chip is introduced, specifically designed for the \textit{OSort} online training algorithm described in \cite{rutishauser2006}. However, it suffers from a large memory requirement and a limited number of channels due to its power consumption. Reference \cite{zeinolabedin2022} performs multi-channel on-chip spike detection, feature extraction, and training and inference for spike sorting with low power consumption. Nonetheless, its capability is restricted to performing Euclidean distance metrics, which shows limitations for some datasets \citep{guo2022}.

Figure~\ref{fig:pwr_analysis} reveals another critical subset of on-chip digital processing engines, that is compression. Unlike spike sorting, compression offers enhanced potential for off-chip signal reconstruction, particularly with lossless compression methods. This capability provides significant advantages, especially in explorative neuroscientific applications, where a visual inspection of the raw signal is required. Additionally, compression finds broader utility across various contexts. For instance, in the context of local field potentials (LFPs) and multi-unit activity (MUA), spike sorting proves inadequate. This point will be further elucidated in the following section, along with the characterization of LFPs. According to \cite{guoliyuan2023}, the compression of action potentials (APs) is categorized into three groups: lossless compression, near-lossless compression, and lossy compression. Reference \cite{bonfanti2010} introduces a compression engine employing a hard thresholding spike detection and Manchester code, achieving a near-lossless space-saving ratio (SSR) of 87.8\%. Lossy compression is often performed in the frequency domain. Reference \cite{nejad2014} utilizes the Walsh-Hadamard transform, followed by a threshold operation to filter out certain coefficients, achieving an SSR of nearly 90\% with a firing rate of approximately 55 to 60 spikes/sec. Similarly, \cite{Shaeri2011} and \cite{kamboh2009} employ discrete wavelet transforms. Reference \cite{thies2019} converts spikes into a feature space to achieve a lower data rate; however, it fails to encode important interspike intervals necessary to recover precise spike timings. These methods achieve high SSR at the cost of introducing compression artefacts. In \cite{guoliyuan2023}, a combined lossless and near-lossless compression engine is proposed; however, its scalability is limited by a power consumption of approximately 17~\textmu W/Ch. Regarding LFPs, \cite{lopez2022, WANG2024, valencia2024, khazaei2020, nurse2016, schmale2013} present various compression algorithms along with temporal and spatial decorrelation methods, yet they suffer from challenges such as high hardware complexity, imperfect reconstruction, or insufficient SSR. 

Another highlight of proposed PSoC is that it integrates an ultra-low power CPU. In settings where electrode movement and tissue reactions are commonplace, as observed in contemporary high-density multi electrode arrays (MEAs) such as \cite{ballini2014, han2013, lopez2016, schueffny2022}, the significance of on-chip programmability and re-training, cannot be overstated. Its role is pivotal in facilitating real-time adaptation, enabling dynamic adjustments to compression engines and spike-sorting classifiers, thereby ensuring sustained high performance in SSR and classification accuracy. To meet these demands, the ultra-low power CPU is complemented by hardware accelerators. Despite the critical nature of this requirement, contemporary state-of-the-art works integrating a CPU are rare. Furthermore, there are notable designs such as \cite{karkare2013, karkare2011, chen2023} that emphasize various on-chip data processing aspects but do not incorporate AFEs, hindering their applicability in implantation scenarios.

Following these design considerations, the PSoC presented, comprises:
\begin{itemize}
    \item A digital signal processing wrapper (DSPW) containing a central bio-signal processing unit (CBPU), a compression engine (CE) consisting of intra- and cross-multi-channel compression engines (16$\times$ ICE and CCE), a spike raster (SR), an adaptive threshold estimator (ATE), and a finite impulse response (FIR) filter.
    \item An on-chip ultra-low power RISC-V core, handling primarily the training process of CCE, the training and inference process of spike sorting, and other processor-driven on-chip operations.
    \item A configurable general purpose multiply-accumulate (MAC) unit which accelerates neural processing such as spike sorting. For general use cases, this module is supplied by SRAM memory. Dimensions and features of the MAC unit can be adapted via the control register block of the CBPU control registers.  
    \item A 68-channel delta-sigma 0.4~V 9-bit analog-to-digital converter (ADC) with corresponding 0.55~V digital wrappers (ADW). The AFE is capable of seamless switching between high-bandwidth and low-bandwidth modes. 
\end{itemize}

The PSoC is implemented using 22nm FDSOI technology. The acquisition side (AFE and ADW) achieves a power consumption of 0.41~\textmu W per channel. Additionally, the IC achieves an SSR of approximately 91\% for APs in the near-lossless mode and an SSR of 64\% for LFPs. The ultra-low power RISC-V core consumes a dynamic power of 5.19~\textmu W/MHz. The MAC unit decreases the power consumption of the software-based on-chip spike sorting by approximately 23.8\%, equivalent to 1.09~\textmu J/spike, achieving an average accuracy of 91.48\% or 94.12\% based on utilized features over datasets published in \cite{quiroga2004}. 

The subsequent sections of this paper are structured as follows. Section \ref{sec:mat_and_met} provides a concise overview of the fundamentals of neural signal processing approaches, detailing the proposed architecture and its constituent components. Section \ref{sec:results} presents the results of the measurements. Section \ref{sec:discussion} delves into a comparative analysis, followed by concluding remarks.

\section{Materials and Methods}\label{sec:mat_and_met}

\subsection{Fundamentals of Neural Signal Processing Approaches}
\label{subsec:fundamental}
This subsection is dedicated to characterizing the extracellular signal and discussing prevalent neural signal processing methodologies. This aims to underscore the advanced integration capabilities of the proposed PSoC, demonstrating its versatility as a platform for capturing extracellular neural signals. 

Brain activities can be categorized into four types based on recording positions: electroencephalography (EEG), electrocorticography (ECoG), extracellular signals, and intracellular signals. Due to their spatial resolution and acquisition complexity, extracellular signals have garnered significant attention \citep{dipalo2017}. This work primarily concentrates on recording and processing extracellular neural signals. Typically, extracellular signals comprise two main components:
\begin{itemize}
    \item Action potentials: APs represent rapid rises and falls of the membrane potential of neurons, which have typical a frequency of 100 Hz - 10 kHz \citep{HODGKIN199025, zeinolabedin2022}. Also referred to as spikes, APs serve as fundamental electrical signals crucial for communication within the nervous system. Their dynamics are integral to deciphering the complex mechanisms governing brain function and behavior.
    \item Local field potentials: LFPs are derived from the lowpass filtering of raw extracellular signals recorded from neural tissue surrounding the electrode, typically within a diameter of approximately 1 mm \citep{buzsaki2012, REY2015106}. These LFPs are pivotal in various applications, including the detection of conditions such as Parkinson's disease and seizures \citep{Rowland2015, wu2013, zeng2023}.
\end{itemize}

In addition to APs and LFPs, extracellular signals include other components such as MUA. While MUA provides valuable information for broader measures like average firing frequency and time-to-first-spike, as demonstrated in studies such as \cite{Lee2008, Flint2013}, it is generally considered background noise in spike-sorting scenarios \citep{buzsaki2012}. Since the methods involved in the analysis of such signal features are highly application-dependent, no dedicated hardware accelerator was considered for this task.

The conventional approach to processing APs acquired through extracellular recording, is to use bandpass-filters, and subsequent spike detection and spike sorting. Compared to spike detection and transmitting entire spikes, spike sorting has the potential to conserve more bandwidth and reduce power consumption. Nevertheless, spike sorting methods encounter a significant limitation: the inability to reconstruct the transmitted signal off-chip. This limitation poses challenges for applications such as neuroscientific research and clinical practice \citep{guoliyuan2023} when compared to traditional neural signal acquisition approaches. Moreover, transmitting raw neural signals offers the potential for algorithmic development in spike detection and sorting. Thus, on-chip compression of neural signals becomes crucial as well, particularly focusing on the lossless or near-lossless compression \citep{guoliyuan2023, ma2021}.

The typical processing steps for LFPs encompass several key stages, including filtering, feature extraction, the computation of biomarkers, and more \citep{jackson2017, olive2022, summerson2015}. Additionally, similar to APs, transmitting raw LFPs off-chip is crucial for off-chip data analysis and algorithm development, underscoring the significance of LFP compression as a meaningful processing step. In contrast to APs, LFPs exhibit clear spatial correlation, necessitating a distinct approach to compression \citep{lopez2022, WANG2024, valencia2024, khazaei2020, nurse2016, schmale2013}.

\subsection{Proposed System Architecture}
\label{subsec:sys_archi}
\begin{figure*}[t]
\centerline{\includegraphics[width=180mm,keepaspectratio]{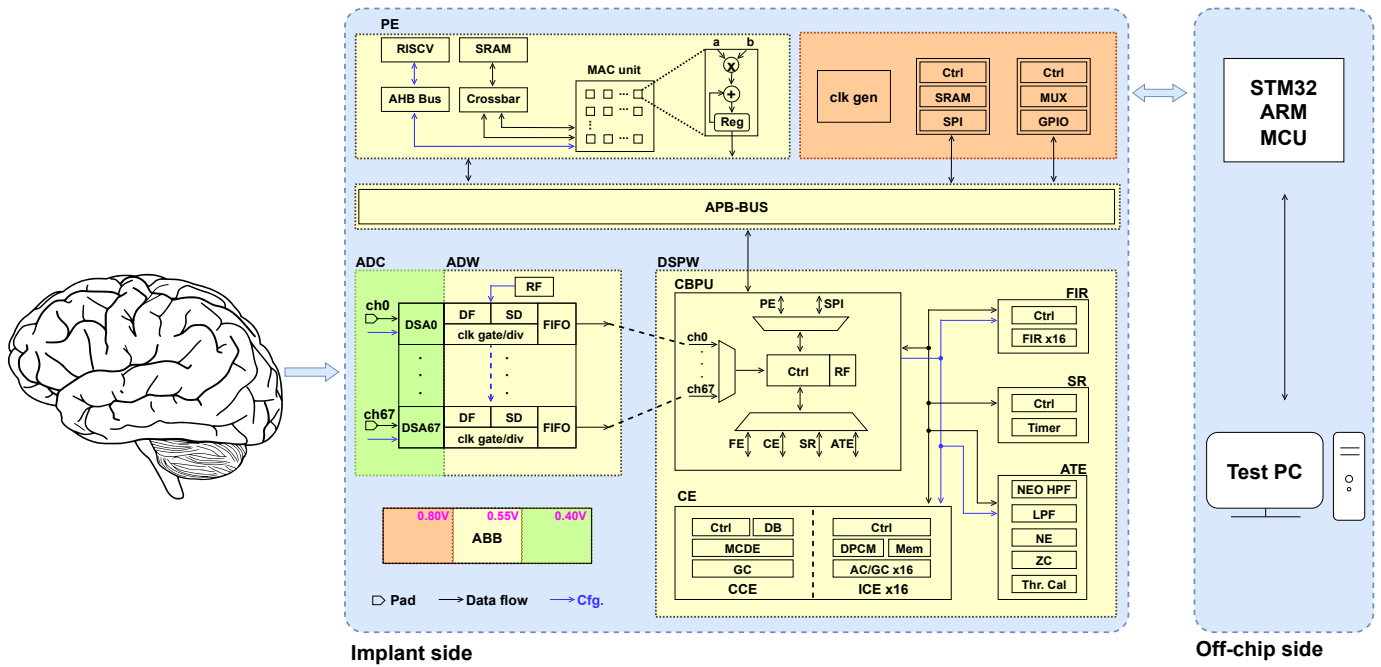}}
\caption{System architecture of proposed PSoC. The PSoC consists of three power domains and the following main components: acquisition analog frontends, digital processing wrapper, MAC-assisted PE, and periphery. }
\label{fig:archi}
\end{figure*}

Figure~\ref{fig:archi} illustrates the system architecture of the proposed PSoC, which is composed of three distinct power domains. The digital processing units, encompassing ADW, DSPW, MAC-assisted PE, and the Advanced Peripheral Bus (APB), are powered at a domain of 0.55~V, facilitated by adaptive body-biasing (ABB) technology, the implementation methodology of which is detailed in~\cite{hoeppner2020}. The delta-sigma modulators of the AFE are robustly powered at 0.4~V, through adaptive back-gate voltage tuning (ABGVT) \citep{schueffny2022} to compensate process-, voltage-, and temperature- (PVT) variation. The third power domain operates at 0.8~V and supplies components such as serial peripheral interface (SPI) and general purpose input/output (GPIO). 

The extracellular neural signal undergoes initial recording through the 68-channel frontends. The configurable ADCs, coupled with digital filters (DF) and spike detectors (SD),  facilitate an effective trade-off between high resolution and ultra-low power consumption. For non-spiking parts, the default low-bandwidth mode is employed. While spikes are detected, the high-bandwidth mode is activated. The two-stage spike detection, encompassing 1) adaptive thresholding and 2) nonlinear energy operator (NEO), ensures accuracy by testing the false positives \citep{schueffny2023}.

Subsequently, the neural signals are temporarily stored in 68 small first-in-first-out (FIFO) buffers located in ADW, under the control of the CBPU. This arrangement aims to maintain a specific order of inputs from all AFEs, thereby ensuring the synchronization of data for potential applications such as spatial decorrelation.

The neural signal is then transmitted to DSPW for the subsequent on-chip processing. The CBPU supports nine distinct commands (C1 - C9), each of which is highly configurable to meet the requirements in various application scenarios. The commands are described briefly as follows:

\begin{itemize}
    \item C1: One of the primary commands involves recording neural signals from AFEs and then transmitting the collected data to PE SRAM, facilitating on-chip data processing capabilities, such as data analysis, and training for CCE and spike sorting. Configurable parameters including channel selection, the frequency of information packets, the address information, and others, are managed through a register file (RF). 
    \item C2: In addition to on-chip data processing, another crucial application involves transmitting neural signals externally for off-chip data processing. Configurable parameters include channel selection, the frequency of information packets, the transmitting mode (stream/batch), and others.
    \item C3: In comparison to transmitting raw neural signals, the transmission of compressed data significantly reduces power consumption and conserves bandwidth. Command C3 is primarily employed for ICE and compression related to AP. Configurable parameters encompass channel selection, CE algorithm selection, compression mode (lossless/near-lossless), transmitting mode (stream/batch), and others.
    \item C4: Command C4 is specifically designed for CCE, primarily utilized for the compression of LFP. Configurable parameters include channel selection, feedback to the Power Management Unit (PMU), transmitting mode (stream/batch), and others.
    \item C5, C6: The commands C5 and C6 are both allocated for utilization within the FIR module. Distinctions arise in their operational outputs: C5 entails transmission of the filtered signals via SPI, whereas C6 involves storage of the signals in the PE SRAM for subsequent on-chip analysis. Additionally, the module serves to filter out AP from raw data, thereby enabling applications such as seizure detection. Configurable parameters encompass channel selection, weights, among others. 
    \item C7: Command C7 is specifically allocated to the spike raster module. Through the integration with the AFE and the two-stage spike detection module, the Spike Raster (SR) and CBPU-ctrl subsystems collaborate to produce timing-accurate spike raster plots. These plots serve a pivotal role in various applications, notably in the reconstruction and analysis of neural circuits.
    \item C8, C9: Commands C8 and C9 are purposefully designed for the adaptive threshold estimator module, serving as the second stage of spike detection. Its primary function revolves around gathering the dynamic thresholds of channels and subsequently transmitting the obtained results off-chip for analysis (C8). Additionally, it facilitates the updating of individual first-stage spike detection parameters (C9).
\end{itemize}

\begin{figure}[tb]
\centerline{\includegraphics[width=85mm,keepaspectratio]{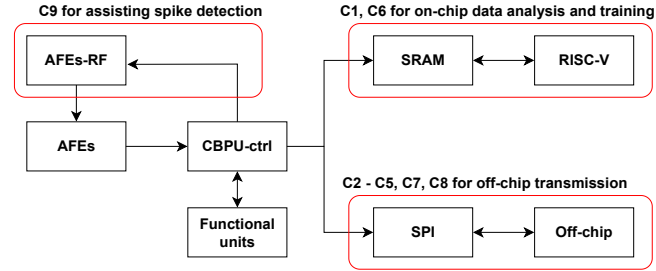}}
\caption{Data flow and active blocks depicting various commands. In addition to the labeled block in the diagram, each command involves the CBPU-ctrl and corresponding functional units. For commands C1 and C6, the process comprises a recording phase followed by an analysis/training phase. During the recording phase, the RISC-V remains in sleep mode to conserve power. }
\label{fig:cmd_block}
\end{figure}

The data flow and corresponding active blocks are illustrated in Figure~\ref{fig:cmd_block}. The commands can be categorized into three types based on their applications: off-chip data transmission, on-chip data analysis and training, and assistance for spike detection.

To verify the functionality of each functional block, a debug mode has also been implemented for commands C2, C3, C4, C5, C7, C8, and C9. In this mode, the CBPU-ctrl retrieves data from the PE SRAM instead of the AFE, allowing users to pre-store data. The debug mode facilitates the debugging process by enabling users to compare the outputs of individual functional modules with the software results obtained off-chip, thereby verifying system performance and aiding in the development of recording protocols. Activation of the debug mode is controlled by a specific bit in the commands.

\subsection{Digital Signal Processing Wrapper}\label{subsec:dspw}
\begin{figure*}[t]
\centerline{\includegraphics[width=180mm,keepaspectratio]{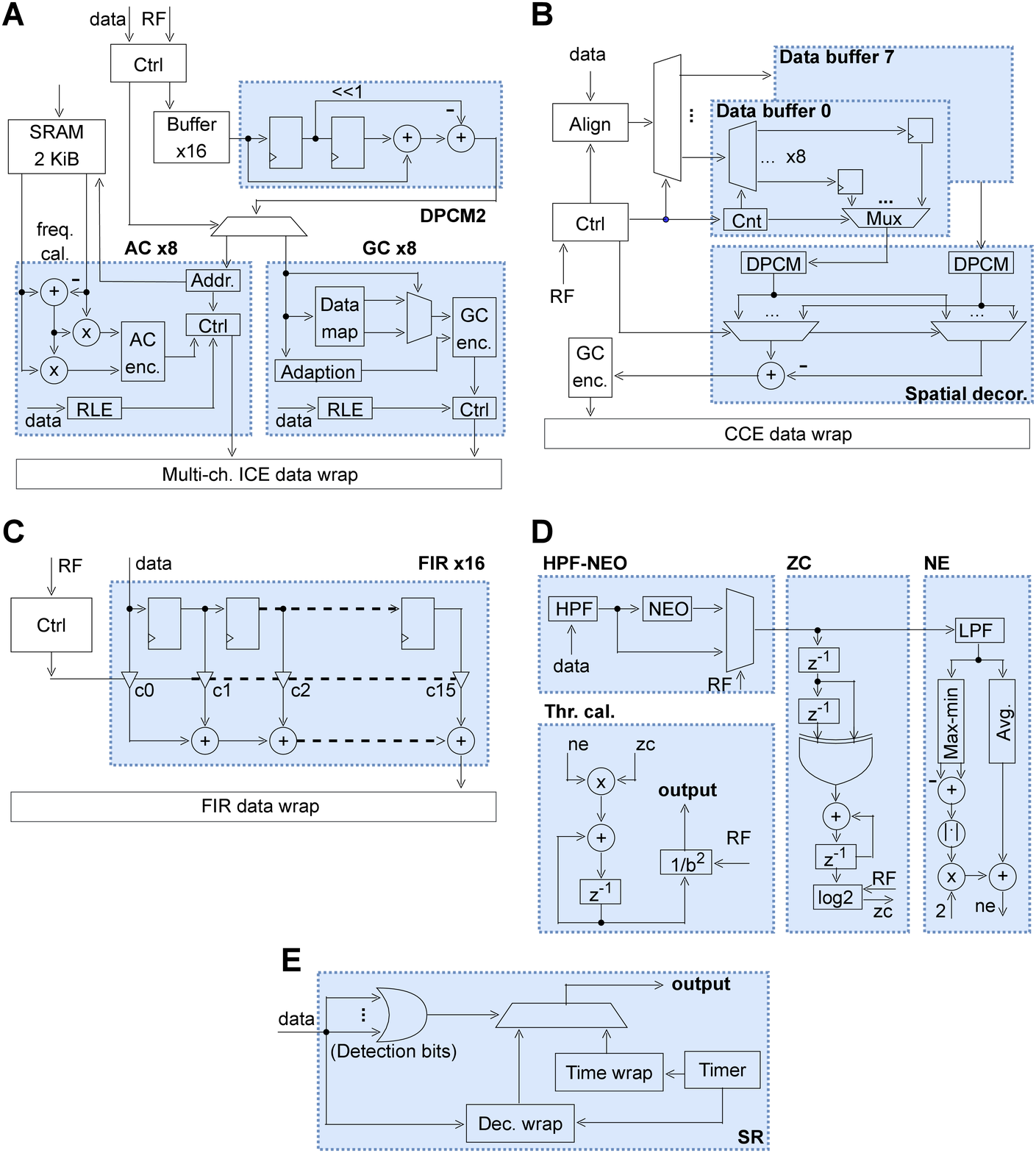}}
\caption{Block diagrams of main components in DSPW. \textbf{(A)} Block diagram of ICE. \textbf{(B)} Block diagram of CCE. \textbf{(C)} Block diagram of the FIR filter. \textbf{(D)} Block diagram of ATE. \textbf{(E)} Block diagram of SR. }
\label{fig:dspw_whole}
\end{figure*}

As illustrated in Figure~\ref{fig:archi}, the proposed PSoC incorporates a versatile DSPW. This subsection provides a comprehensive explanation of the primary functional components comprising the DSPW, the block diagrams of which are shown in Figure~\ref{fig:dspw_whole}. These components fulfill crucial roles in facilitating diverse on-chip extracellular signal processing tasks.

\subsubsection{ICE for AP Compression}\label{subsubsec:ap_ce}
As highlighted in Subsection \ref{subsec:sys_archi}, the CBPU-ctrl unit initially gathers data from either AFEs in normal mode or the PE SRAM in debug mode. Subsequently, it forwards the data from specific user-selected channels to the ICE module for executing lossless or near-lossless compression. In prior research \citep{guoliyuan2023}, an on-chip compression engine was developed, focusing on achieving high SSR, and utilizing adaptive arithmetic coding. Detailed power and area analysis reveals that approximately 75\% of power consumption and 95\% of area are occupied by the adaption module, opening a perspective for further system-optimization.

Figure~\ref{fig:dspw_whole}\textbf{A}
illustrates the block diagram detailing the proposed 
ICE. This engine comprises 16 compression modules, with eight employing AC and the remaining eight utilizing GC. Notably, the system offers flexible configuration options, allowing each of the 68 AFEs to be designated for compression using either AC or GC. Upon receiving data tagged with the designated channel index, it undergoes initial storage in a buffer with a capacity of 32. The underlying concept here is crucial for near-lossless compression, where upon detecting a spike, up to 32 preceding samples are considered part of the current spike, ensuring precise spike reconstruction. The subsequent step involves DPCM2, a straightforward yet effective time-domain decorrelation technique. Compared to the DPCM method outlined in \cite{guoliyuan2023}, DPCM2 notably concentrates the raw data distribution, a critical aspect for enhancing the performance of entropy encoding algorithms such as AC and GC.

Initially, we delve into how the AC engine enhances area efficiency and reduces power consumption while incurring a minimal decrease in SSR. Unlike adaptive arithmetic coding, AC eliminates the need for an adaptive module. Instead, it employs a 2 KiB SRAM to store the distribution of all symbols, shared among eight AC engines. This distribution is established through training. Experimental findings indicate that multiple channels can share similar distributions, conserving memory space while ensuring each channel maintains a high SSR. This approach embodies a form of semi-adaptive compression engine.

In contrast to AC, the GC engine exhibits a slightly lower SSR; however, it compensates with unparalleled efficiency in both area and power consumption. Following decorrelation via DPCM2, the subsequent stage for GC involves an efficient data mapping process. The reasons are that 1) GC requires non-negative inputs, and 2) the smaller the inputs, the higher the SSR. The output from the data mapping module is then directed to the GC encoder. To optimize GC performance, an adaptation module is incorporated, primarily responsible for tracking the proportion of zero samples. 

In the near-lossless module, all samples between spikes are treated as zeros, and a run-length encoding (RLE) module is incorporated to quantify the length of these zero intervals. The RLE output is subsequently split into 2 or 3 components bitwise and transmitted to either Arithmetic Coding (AC) or Golomb Coding (GC) modules as per the chosen encoding scheme. During decoding, data is grouped into sets of 66 or 67 samples. The initial 2 or 3 samples within each group are utilized to derive the RLE result, succeeded by a spike comprising 64 samples, denoting the end of the zero interval and the resumption of non-zero data.

In our ICE, we explored the implementation of both AC and GC to demonstrate their compatibility within a multi-channel AP compression engine. GC offers notable advantages in terms of area and power efficiency, rendering it particularly advantageous in scenarios with an extensive number of channels. However, in instances of highly noisy data, its SSR tends to be noticeably lower than that of AC. This aspect gains significance, especially in applications involving wireless transmission, as further discussed in Section \ref{sec:intro}. 

\subsubsection{CCE for LFP Compression}\label{subsubsec:lfp_ce}

Compared to multi-channel action potential signals, LFP signals exhibit a more pronounced correlation in the spatial dimension. This highlights the insufficiency of relying solely on temporal decorrelation to achieve a high SSR. Therefore, the primary objective of LFP compression is to efficiently decorrelate signals in both the temporal and spatial dimensions. 

Figure~\ref{fig:dspw_whole}\textbf{B} 
shows the block diagram of an 8-channel CCE designed for LFP compression. As explained, spatial decorrelation holds significant importance in LFP compression, necessitating precise alignment of data from various channels. This alignment process unfolds in two stages. Initially, CBPU-ctrl undertakes the responsibility of aligning data from different AFEs to ensure synchronization. Subsequently, within the CCE, the align module initiates the operation of data buffers upon the delivery of data from the channel with the smallest index among the selected eight channels. All eight data buffers used in CCE are tiny compared to the buffers used for ICE, since we do not need to care about buffering entire spikes. 

Except for the root channel obtained during the training step, each channel has a single parent channel. This implies that the samples from the parent channel will be utilized for spatial decorrelation for the child channel. The basic formula of this process is defined as: 
\begin{equation}\label{eq:tab1}
    \tilde{e} _{c} (n)=e_{c}(n)-\gamma \cdot e_{r}(n)
\end{equation}
where the suffix \textit{c} denotes the child channel, the suffix \textit{r} signifies the reference channel (parent channel). Additionally, $e_c$ and $e_r$ represent the temporally decorrelated signals of child and parent channels, respectively. The symbol $\gamma$ denotes the spatial decorrelation factor, and $\tilde{e}_{c}$ is the temporally and spatially decorrelated signal, which serves as the data to be compressed for the corresponding child channel.
To achieve the optimal decorrelation outcome, precise calculation of $\gamma$ is essential. As per \cite{kamamoto2008}, there exists a positive correlation between the final coding length and energy, which in this context is defined as:
\begin{equation}\label{eq:energy}
    \tilde{E}_{c}=\sum_{n=1}^{N}(\tilde{e} _{c} (n))^2
\end{equation}
where $N$ denotes the window size utilized for computing $\gamma$. By taking the derivative of energy with respect to $\gamma$ and solving for the point at which the derivative equals zero, we obtain:
\begin{equation}\label{eq:gamma}
\gamma = \frac{\sum_{n=n_0}^{N}e_{c}(n) \cdot e_{r}(n)}{\sum_{n=n_0}^{N}e_{r}(n) \cdot e_{r}(n)}
\end{equation}
where $n_0$ indicates the starting point for $\gamma$ calculation.

Equation (\ref{eq:tab1}) is commonly referred to as the one-tap mode. Alternatives include the three-taps mode or five-taps mode. The primary distinction lies in the number of values derived from the reference channel. These alternatives are also investigated in our experiments using datasets from \cite{WATSON2016, vandecasteele2012}. However, despite the evident increase in computational complexity with multi-taps mode, there is hardly any improvement observed in SSR.

The CCE requires a training process to establish the chain of channels and to calculate $\gamma$. As demonstrated in our experiments, we have verified that when the window size ranges from 1000 to 2000, the $\gamma$ values of channels stabilize significantly. This indicates that on-chip training encounters no issues. Furthermore, a minimum spanning tree is utilized in the training process to ensure that the chain of channels does not form a closed-loop.

\subsubsection{FIR Filter}\label{subsubsec:fe}
For processing particular features of the neural signal e.g. LFPs, a discrete-time FIR filter is implemented based on the following formula:
\begin{equation}\label{eq:FIR}
\begin{aligned}
    y\lbrack n \rbrack &= c_0 \cdot x \lbrack n \rbrack + c_1 \cdot x \lbrack n - 1 \rbrack + \cdots + c_N \cdot x \lbrack n - N \rbrack \\  &= \sum_{i=0}^N c_i \cdot x \lbrack n - i \rbrack
\end{aligned}
\end{equation}

The implemented functionality can be utilized in concert with the PE, in order to efficiently implement the large group of commonly used algorithms that are based on convolutions. Among these, various transforms can be applied to the recorded discrete time series. For example, Discrete Wavelet Transforms, formulated as dyadic filter banks, can be employed to extract spectral features for event detection. 

The CBPU-ctrl unit receives data from the AFEs and forwards the data to the filter module whenever new data arrives. The filter module is capable of dealing with data originating from up to 16 channels in parallel. It consists of a wrapper that instantiates 16 FIR taps, each of which has a filter length of 16. Each FIR tap is of order 15 and receives its data from the FIR data wrap module. 
The FIR data wrap maps the data from each of the 16 channels to one of the FIR taps. The filter coefficients (denoted as \(c_0 \) to \( c_{15}\) in 
Figure~\ref{fig:dspw_whole}\textbf{C}) 
are set in the register file and have a word width of 16 bits. Each FIR tap has an accumulator  resolution of 26 bits (denoted as plus-sign in Figure~\ref{fig:dspw_whole}\textbf{C}). 
The output of each FIR tap is sent to the FIR data wrap which itself is mapping the FIR tap output to the appropriate channel index. The detailed block diagram of one FIR tap is shown in Figure~\ref{fig:dspw_whole}\textbf{C}. 

\subsubsection{Adaptive Threshold Estimator}\label{subsubsec:ate}
Figure~\ref{fig:dspw_whole}\textbf{D} 
illustrates the block diagram of the adaptive threshold estimator, comprising primarily a NEO module, a zero crossings (ZC) module, a noise estimator (NE) module, and a threshold calculator module.

The primary purpose of introducing the NEO module, as defined by Equation (\ref{eq:neo}), is to enhance the detection of spike activity amidst background noise. Typically, spikes exhibit sudden changes in amplitude compared to noise.
\begin{equation}\label{eq:neo}
    \Psi [x(n)]=x(n)^2 - x(n-1)\cdot x(n+1)
\end{equation}
where $\Psi [X(n)]$ denotes the output of the NEO module corresponding to the neural signal $x(n)$. As per \cite{zeinolabedin2022, mukhopadhyay1998}, the NEO enhances the signal-to-noise ratio of the signal, thereby reducing its sensitivity to a threshold value, as observed in scenarios lacking NEO \citep{obeid2004, gibson2012, zeinolabedin2015}. 

The ZC module serves to tally the zero crossings in the output of the NEO module, subsequently utilizing this count to estimate the firing rate. When consecutive points possess different sign bits, the counter within the ZC module increments by one. To more accurately represent the spike rate in hardware, a logarithmic operation is employed to compress the window size.

The NE module primarily estimates the noise level present in the output of the NEO module. Initially, a low-pass filter within this module is applied to eliminate spikes, facilitating a more precise assessment of the noise level.

Subsequently, the threshold calculator module amalgamates the outcomes of the ZC and NE modules to compute the corresponding threshold. This approach comprehensively accounts for the impact of spike rate and background noise on the threshold. By enhancing spike detection accuracy, it dynamically adjusts the thresholds of the AFE and SD modules, thereby optimizing the power consumption of the frontend acquisition.

The ATE is shared among 68 channels and operates independently for the single channel selected at any given time. Consequently, a working frequency of 20~kHz proves adequate, resulting in low power consumption of the ATE.

\subsubsection{Spike Raster}\label{subsubsec:sr}
The spike raster module is tasked with producing spike raster plots, which visualize the spiking patterns of a neural ensemble across time.

The block diagram of the SR system is depicted in Figure~\ref{fig:dspw_whole}\textbf{E}.
Upon initiating SR with the respective command, the CBPU-ctrl continuously monitors the detection bits of each sample obtained from all 68 AFEs, following the sequence from channel index 0 to 67. Subsequently, the observed results are relayed to the SR module. Within the SR module, an OR-gate operation is conducted across the 68 detection bits. If no spikes are detected in any channel, an empty packet with a designated packet header is transmitted to the CBPU-ctrl. Conversely, in the event of spike detection, a packet indicating the firing channels along with timing details is dispatched to the CBPU-ctrl. Within the CBPU-ctrl, packets received from the SR are transmitted at an approximate frequency of 20 kHz. This implies that upon completion of a round of checking across all 68 AFEs by the CBPU-ctrl, a packet is constructed and conveyed via SPI.

\subsection{MAC-Assisted Processing Elements}
\label{subsec:mac_pe}

\subsubsection{Processing Elements}\label{subsubsec:pe}

\begin{figure}[t]
\centerline{\includegraphics[width=85mm,keepaspectratio]{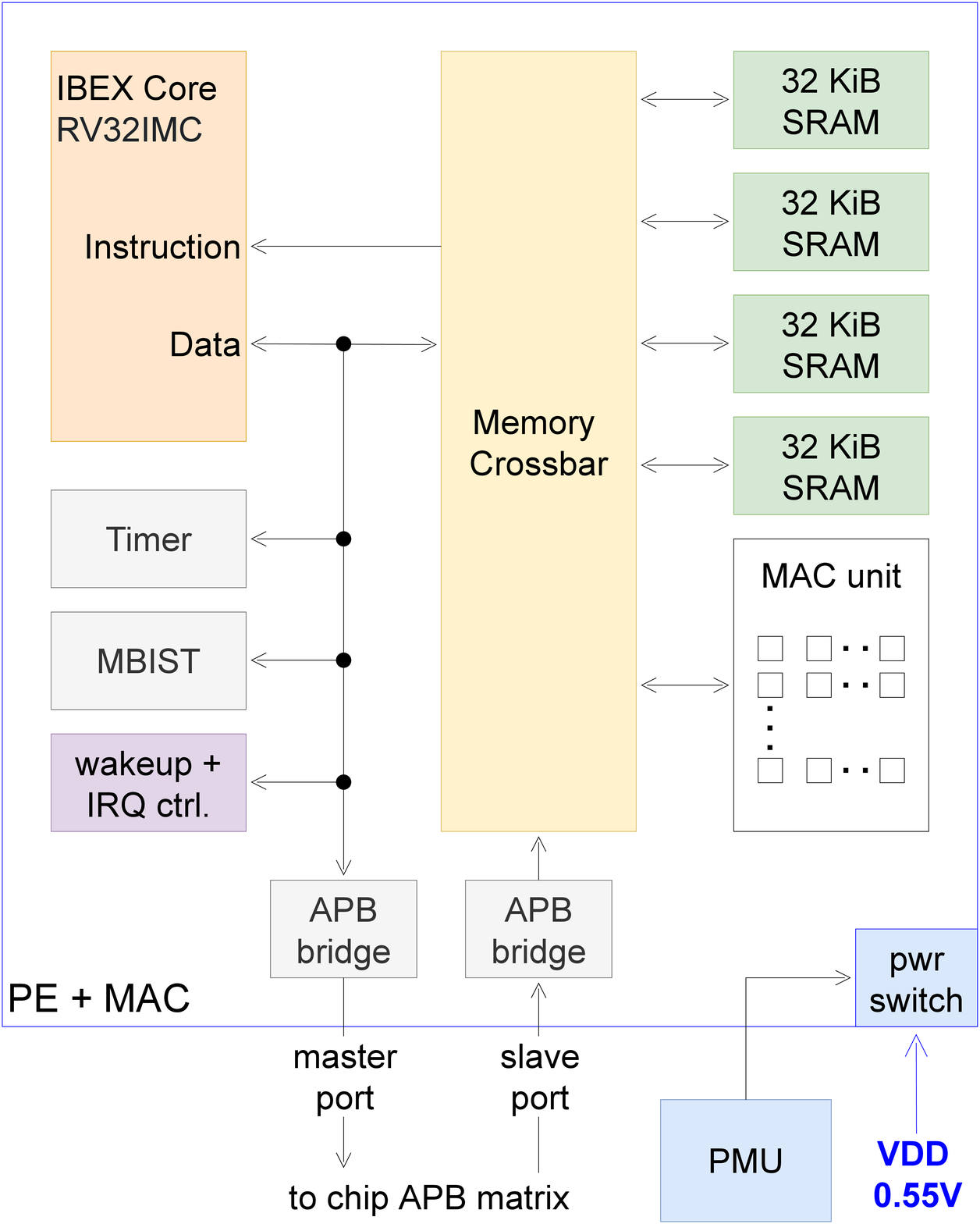}}
\caption{Block diagram of PE and MAC.}
\label{fig:PE_arch}
\end{figure}

Our RISC-V PE architecture is shown in Figure~\ref{fig:PE_arch} and was adapted from \cite{bauer2023}. The main design and implementation goal was to reduce static- as well as dynamic- power consumption. For this purpose, an adaptive reverse body biasing scheme \citep{walter2020} was utilized at 0.55~V supply voltage. Moreover, power switches are implemented, to power-gate the PE if not needed. The used IBEX RISC-V core supports RV32IMC instructions and achieves an architectural performance of 3.12~CoreMarks/MHz. To reduce leakage power, the PE is implemented with a target frequency of 25~MHz using only regular-Vt standard cells. A 128~KiB SRAM organized in four banks of 32~KiB SRAM can be used flexibly as instruction and data memory. The SRAM can be accessed by the PE and MAC, as well as other chip components via the APB bus, allowing various data flow scenarios. To test SRAM memory after production, a Memory Built-In Self Test (MBIST) block was implemented. A flexible wake-up and interrupt controller allows to clock-gate internal components while the PE is sleeping. To further reduce static power consumption, a retention sleep mode was implemented. This allows to power down SRAM periphery while retaining memory contents. Various interrupt sources (timer, MAC, APB access and external IRQ lines) can be used to wake up the PE from its sleep modes.

\subsubsection{MAC Unit}\label{subsubsec:mac}
\begin{figure}[t]
\centerline{\includegraphics[width=85mm,keepaspectratio]{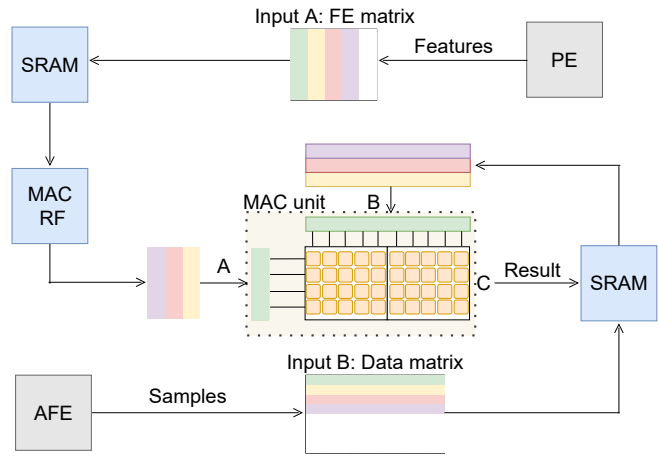}}
\caption{Dataflow of the MAC unit. The FE matrix is stored in SRAM or the internal register file depending on the working flow. Upon being triggered by the PE, the MAC unit executes the required multiply-accumulate operations in the specified order (from green to purple) and then writes the results (C) back into the SRAM.}
\label{fig:MAC_Dataflow}
\end{figure}

The MAC unit is designed to support various neural signal processing applications. Though the MAC unit can work independently, it functions as a support module for the RISC-V CPU. Therefore it requests and sends data from SRAM as shown in Figure~\ref{fig:MAC_Dataflow}. To scale down memory requests, the MAC unit is caching its feature matrix from SRAM into internal register files, as presented in Figure~\ref{fig:MAC_Dataflow}.  This reduces time and power consuming memory requests in inference mode, when utilized for spike sorting. Because only samples are requested from SRAM in this case. To ensure high accuracy for the processing the accumulator of the MAC unit has a bit size of 32. The MAC unit consists of multiple cells processing 9x16-bit signed operations in parallel.  To increase energy efficiency the PE can be set into sleep mode during the execution of the MAC unit. The control, input- and output-dimensions and features of the MAC unit are accessible through registers. This way the whole processing chain of neural signal processing is efficiently controlled and synchronized with each other.   

\subsubsection{Spike Sorting}\label{subsubsec:SS}
\begin{figure}[t]
\centerline{\includegraphics[width=85mm,keepaspectratio]{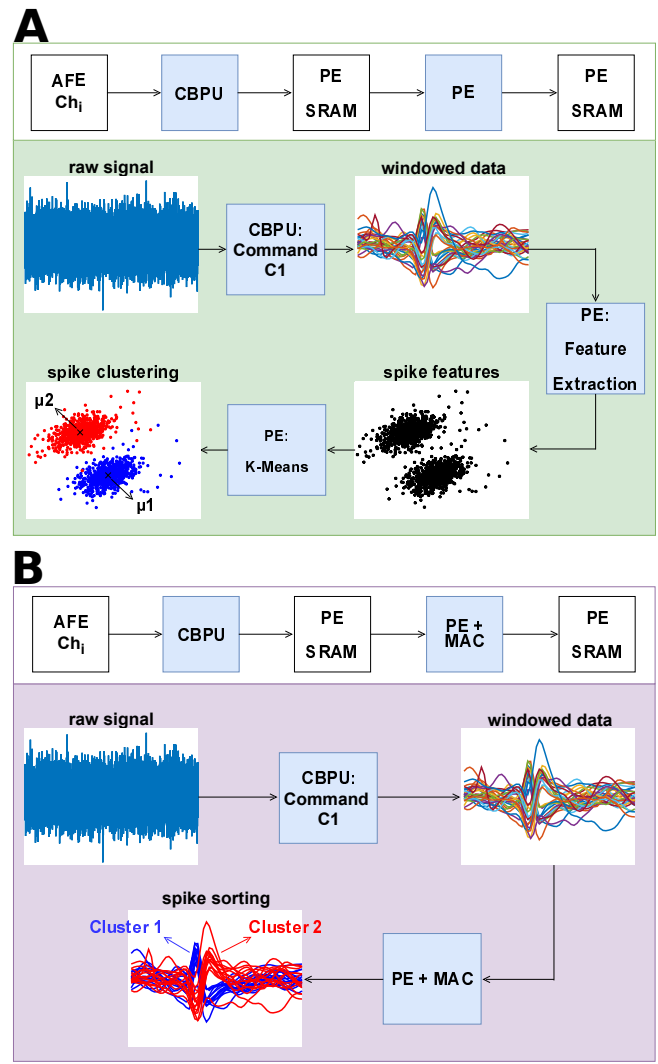}}
\caption{Illustration and process of spike sorting. \textbf{(A)} The training process of the neural spike sorting in PE flow. The neural signals are recorded from AFEs and stored in SRAM memory. In this instance, the PE computes spike features via PCA and partitions the occurring spikes into clusters via k-means clustering. \textbf{(B)} The inference process of the neural spike sorting in PE and MAC flow. The usage of the MAC unit improves the energy efficiency of spike sorting by around 23.8\% compared to performing these spike sorting steps in PE alone.}
\label{fig:ss_flow_whole}
\end{figure}

As introduced, spike sorting plays a crucial role in processing AP signals, the training and inference flows of which are shown in Figure~\ref{fig:ss_flow_whole}. The MAC unit here serves to accelerate the spike sorting algorithm, supporting various feature extraction methods such as principal component analysis (PCA) and adaptive filter (AF), offering hence increased flexibility for on-chip spike sorting. It receives data from SRAM memory and can be deployed in different feature extraction methods, in which multiply-accumulate-operations are the basis. For instance, in this work, PCA in combination with K-means was selected to demonstrate feature extraction in the neural spike sorting and clustering algorithm, as depicted in Figure~\ref{fig:ss_flow_whole}\textbf{A}. 
During the training mode, the principal components of the neural signal are stored in SRAM. 
Figure\ref{fig:ss_flow_whole}\textbf{B} 
illustrates the inference flow of spike sorting. In this process, the MAC unit conducts FE on spikes utilizing the training results, aiding in the spike sorting procedure. The distance calculation in the spike sorting process is carried out by the PE.

\begin{figure}[t]
\centerline{\includegraphics[width=85mm,keepaspectratio]{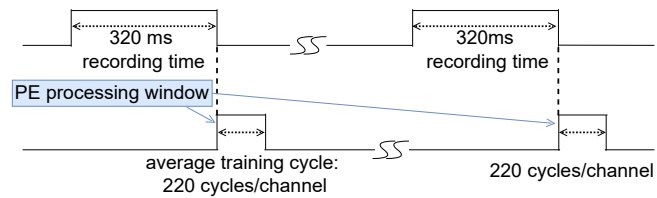}}
\caption{Training time analysis.}
\label{fig:Timing}
\end{figure}

PCA proves to be an efficient feature extraction method for atypical datasets, streamlining feature selection during the training process \citep{GLASER1968, abeles1977}. However, in most scenarios, AF emerges as a more efficient and straightforward feature extraction method \citep{zeinolabedin2022}. Therefore, AF is generally chosen as the feature extraction method in our work. Using AF, it takes an average of 220 clock cycles per channel to process an input sample, as illustrated in Figure~\ref{fig:Timing}. During the recording phase and MAC operating phase, the PE remains in sleep mode. In inference mode, the MAC unit computes the projection of recorded spike samples into the spike feature space. During this phase, only sampled spikes are transmitted to SRAM, fetched by the MAC unit, and entered onto the predefined and already cached FE matrix. 

The advantage of software-based spike sorting extends beyond flexibility in feature extraction to include the inference process as well. While most spike sorting hardware engines solely support
Euclidean distance metrics, which may be insufficient for certain datasets \citep{guo2022}, the software-based spike sorting in this study can accommodate complex distance metrics such as Mahalanobis distance metrics.

\subsection{Recording Frontends}\label{subsec:afe}

\begin{figure}[t]
\centerline{\includegraphics[width=85mm,keepaspectratio]{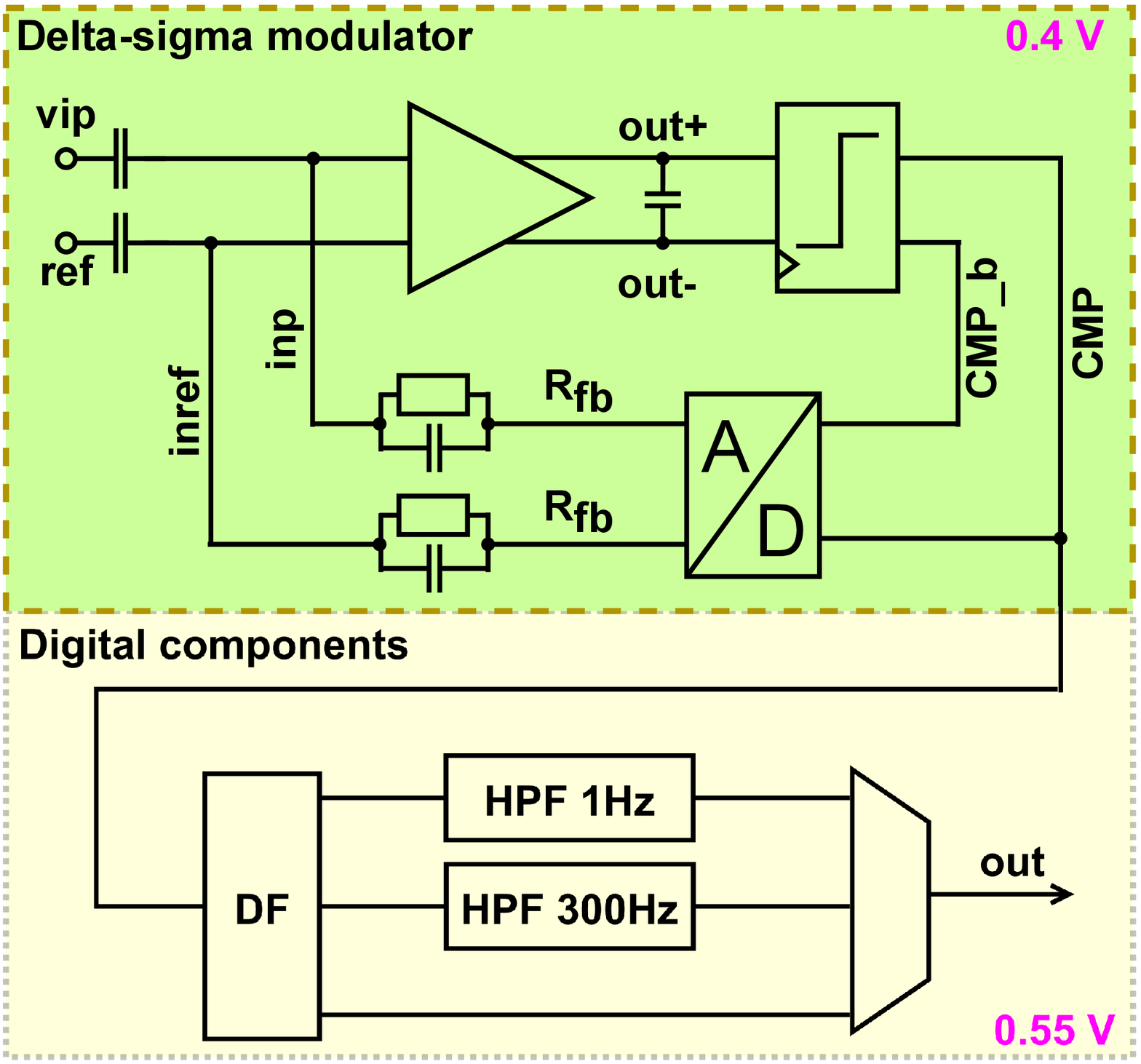}}
\caption{Architecture of analog frontend.}
\label{fig:afe1}
\end{figure}

Figure~\ref{fig:afe1} shows the architecture of the recording frontend. AP and/or LFP are sampled through 68 continuous-time delta-sigma ADCs with corresponding digital high- and low-pass filter acting as decimation filters. The latter can be set to high bandwidth- (10~kHz) or low bandwidth-mode (2.4~kHz). The delta-sigma modulator runs at 0.4~V, by applying forward biasing at the back-gates \citep{schueffny2023}. Its current and sampling frequency is reduced in low-bandwidth mode. The feedback is a parallel RC network to form a high-pass filter with the input decoupling capacitors. The resistor is implemented as a pseudo resistor, which defines the DC voltages at the input of the integrator. The output of the integrator is digitized with a 1~bit comparator. Thus the digital-to-analog converter to close the loop has the same resolution. The sampling frequency in low-bandwidth mode is 1.25~MHz and 5~MHz in high-bandwidth mode. Since low-bandwidth mode is sufficient to record spikes, it is used in this paper. The implemented high-pass filter can be set to 300~Hz to remove local field potential, 1~Hz to remove offset only or be disabled for characterization purposes. Implementing the filters digitally, increases PVT robustness significantly, in comparison to analog filter implementations. To save power, the first stage of the decimation filter is a CIC filter with an asynchronous implemented integrator. The second stage is a polyphase FIR filter with a decimation of two followed by another polyphase FIR filter with a decimation of two. The asynchronous integrator as well as the polyphase approach reduce switching and thus power dynamic current in the circuit. The digital components are supplied by 0.55~V. The delta-sigma modulators are combined with a digital-on-top approach combined with the digital filters to the ADC macro. This macro is instantiated 68 times on toplevel enabling further scaling for more channels. 
Details about a similar analog frontend are presented in \cite{schueffny2022,schueffny2023}.

Regardless of the recording previous publications, show the capability of implementing neural stimulators in 22nm FDSOI \citep{schueffnystim}, enabling a closed-loop system on a chip.

\section{Results}\label{sec:results}
\begin{figure}[t]
\centering
\centerline{\includegraphics[width=85mm,keepaspectratio]{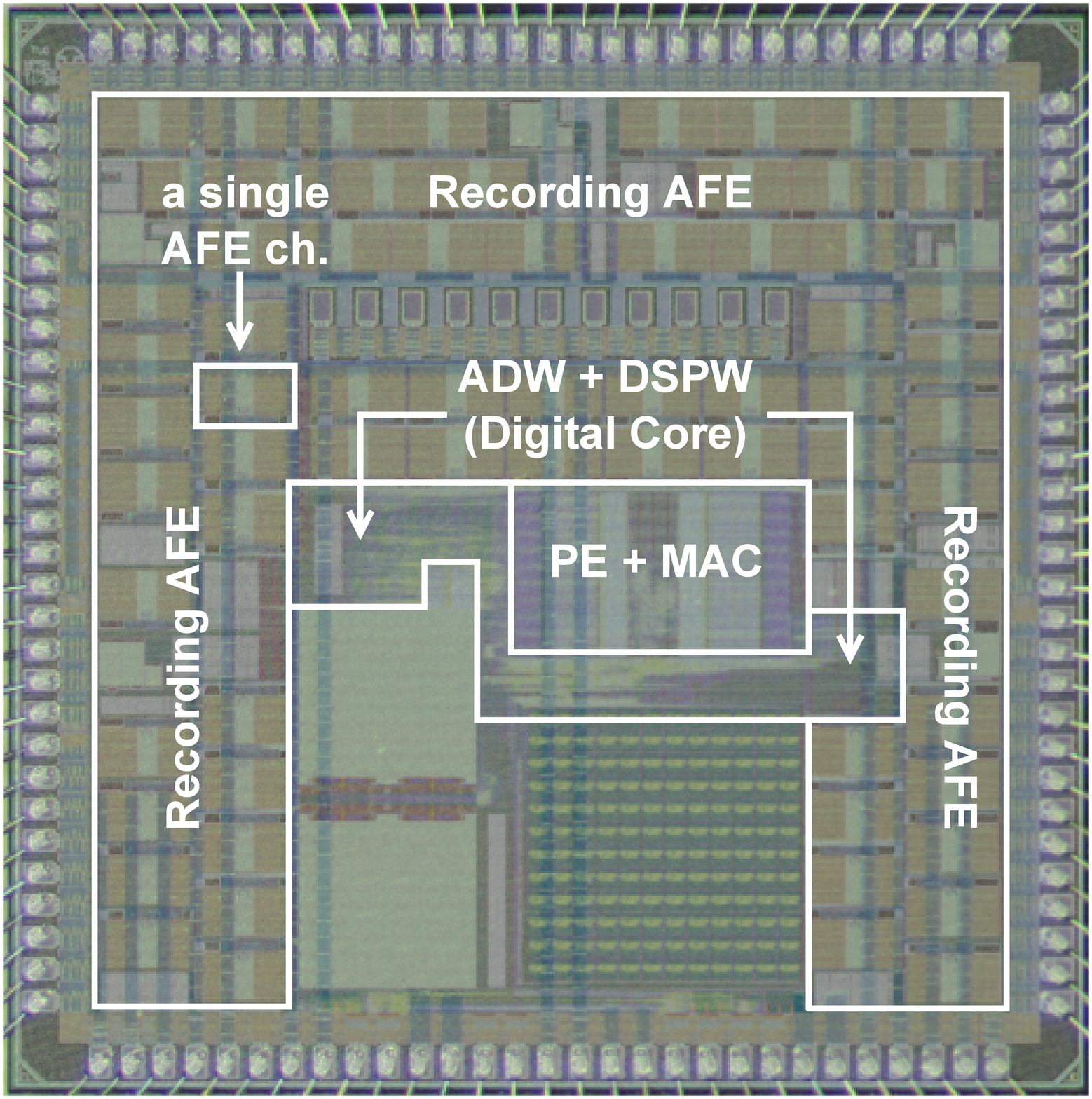}}
\caption{Chip photo, 3mm$\times$3mm, components marked, and transparent layout overlaid.}
\label{fig:die}
\end{figure}

The proposed neural signal recording and processing system has been implemented using 22nm GlobalFoundries FDSOI technology. Figure~\ref{fig:die} displays the chip photo, annotated with components and overlaid with the transparent layout. This section presents power and area measurements, as well as a comparison with other relevant works.

\subsection{Setup for Measurement and Validation}\label{subsec:setup}
\begin{figure}[t]
\centerline{\includegraphics[width=85mm,keepaspectratio]{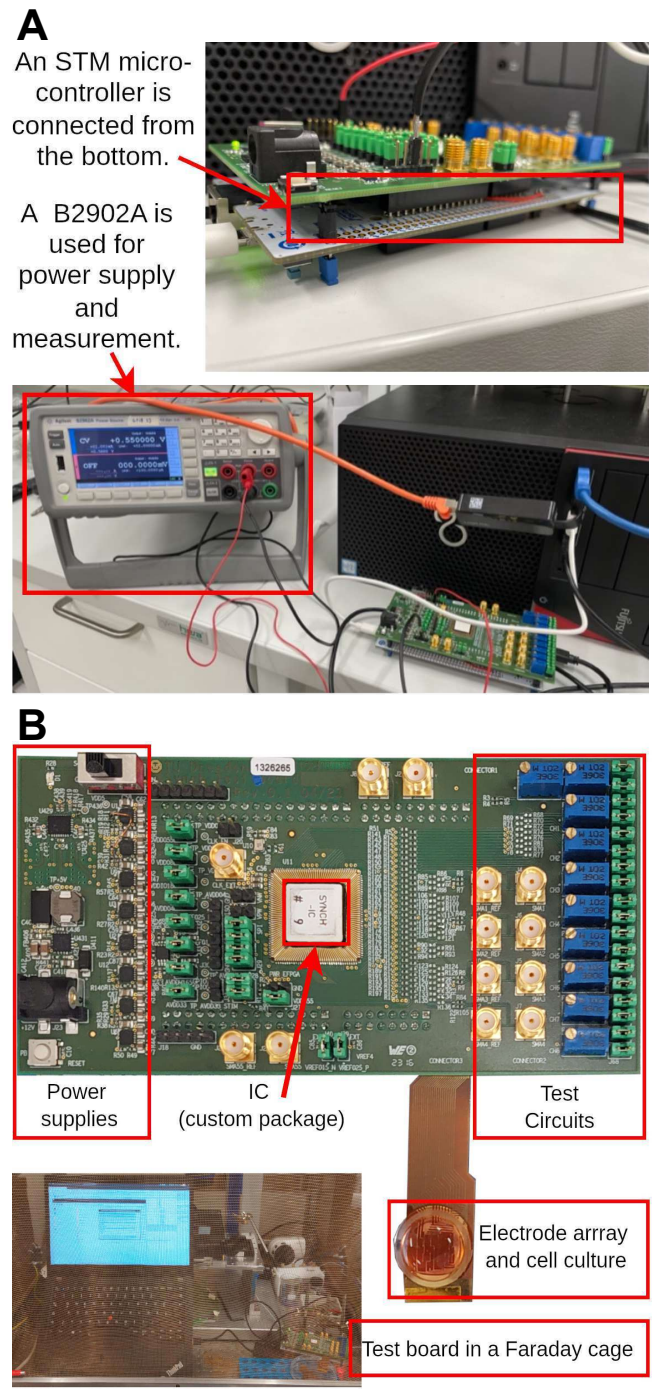}}
\caption{Setup for chip test, measurement and in-vitro experiments. \textbf{(A)} Setup for chip power measurement. \textbf{(B)} Top view of the chip along with PCB board, electrode array and cell culture, accompanied by a photograph depicting the recording environment utilizing our chip.}
\label{fig:setup_whole}
\end{figure}

The different operating modes and functional units were tested as depicted in Figure~\ref{fig:setup_whole}\textbf{A}. 
An STM microcontroller is connected at the bottom of the PCB board, serving as a bridge for data communication between the IC and a computer. During testing, datasets are initially read and converted to analog signals in the testbench to serve as inputs. To evaluate the power consumption, a B2902A precision lab power supply was used to supply the VDD055 domain. All measurements were carried out at room temperature of 20\textcelsius.

The chip is connected to an electrode array and cell culture to enable the actual recording of neuron activities in vitro. To safeguard against external electromagnetic interference, a Faraday cage is employed to shield the chip, its associated electronics, and the electrodes, as shown in Figure~\ref{fig:setup_whole}\textbf{B}.

\subsection{Power and Area}\label{subsec:pow_area}
As emphasized, the efficiency in power consumption and area of PSoC holds paramount importance in assessing the practicality and viability of systems within the neuroprosthetics domain. Hence, we undertake a comprehensive characterization of the digital domain of the proposed PSoC with regard to power and area in this subsection. The power consumption of all digital processing units is measured at 0.55~V supply voltage by applying the ABB technique.

\begin{figure}[t]
\centerline{\includegraphics[width=85mm,keepaspectratio]{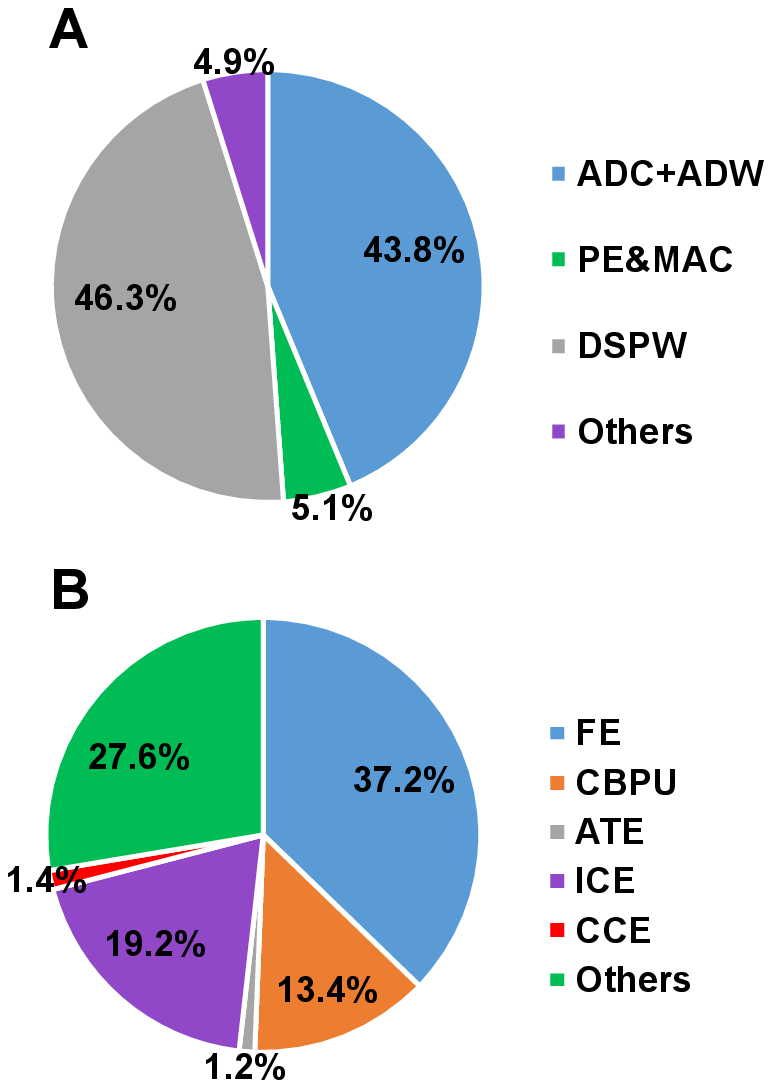}}
\caption{Leakage power break-down to characterize the PSoC. \textbf{(A)} The leakage power break-down of the proposed PSoC. \textbf{(B)} The leakage power break-down of the DSPW block.}
\label{fig:lp_breakdown_whole}
\end{figure}

The ICE comprises two distinct compression engines utilizing AC and GC techniques, respectively. Generally, AC requires a maximum of 120 clock cycles to compress a single symbol, whereas GC requires significantly fewer clock cycles. Given an acquisition frequency of 20~kHz, ICE necessitates a minimum frequency of 2.4~MHz. However, the near-lossless mode imposes a more stringent timing constraint. As elucidated in Subsection \ref{subsec:dspw}, in the worst-case scenario of the near-lossless mode, 64 samples must be compressed within a recording time of 32 samples. Hence, to ensure real-time processing of neural signals, a frequency of 5~MHz is required.
Under this condition, each AC exhibits an average dynamic power of 3.50~\textmu W/Ch coupled with a leakage power of 0.89~\textmu W/Ch in the lossless mode. In the near-lossless mode, the dynamic power is reduced to 3.24~\textmu W/Ch. In contrast, the GC engine demonstrates greater power efficiency, albeit with a slight drop in SSR. In the lossless mode, each channel incurs an average dynamic power of 0.54~\textmu W/Ch and a leakage power of 0.33~\textmu W/Ch, while in the near-lossless mode, the dynamic power is measured at 0.42~\textmu W/Ch.
For CCE, despite the fewer clock cycles required by GC, a single GC engine is shared among 8 channels. Hence, a frequency of 5~MHz is maintained. The 8-channel CCE consumes a dynamic power of 1.86~\textmu W and a leakage power of 0.70~\textmu W, totaling 0.32~\textmu W/Ch. 
The FIR filter operates across 16 channels in parallel. To meet real-time processing demands, a frequency of 50~MHz is selected. The average total power consumption of the 16-channel FIR filter is measured at 25.76~\textmu W. 

The leakage power break-down of the proposed PSoC is shown in Figure~\ref{fig:lp_breakdown_whole}\textbf{A} 
and the leakage power of the digital signal processing wrapper is shown in Figure~\ref{fig:lp_breakdown_whole}\textbf{B}. 

\begin{figure}[t]
\centerline{\includegraphics[width=85mm,keepaspectratio]{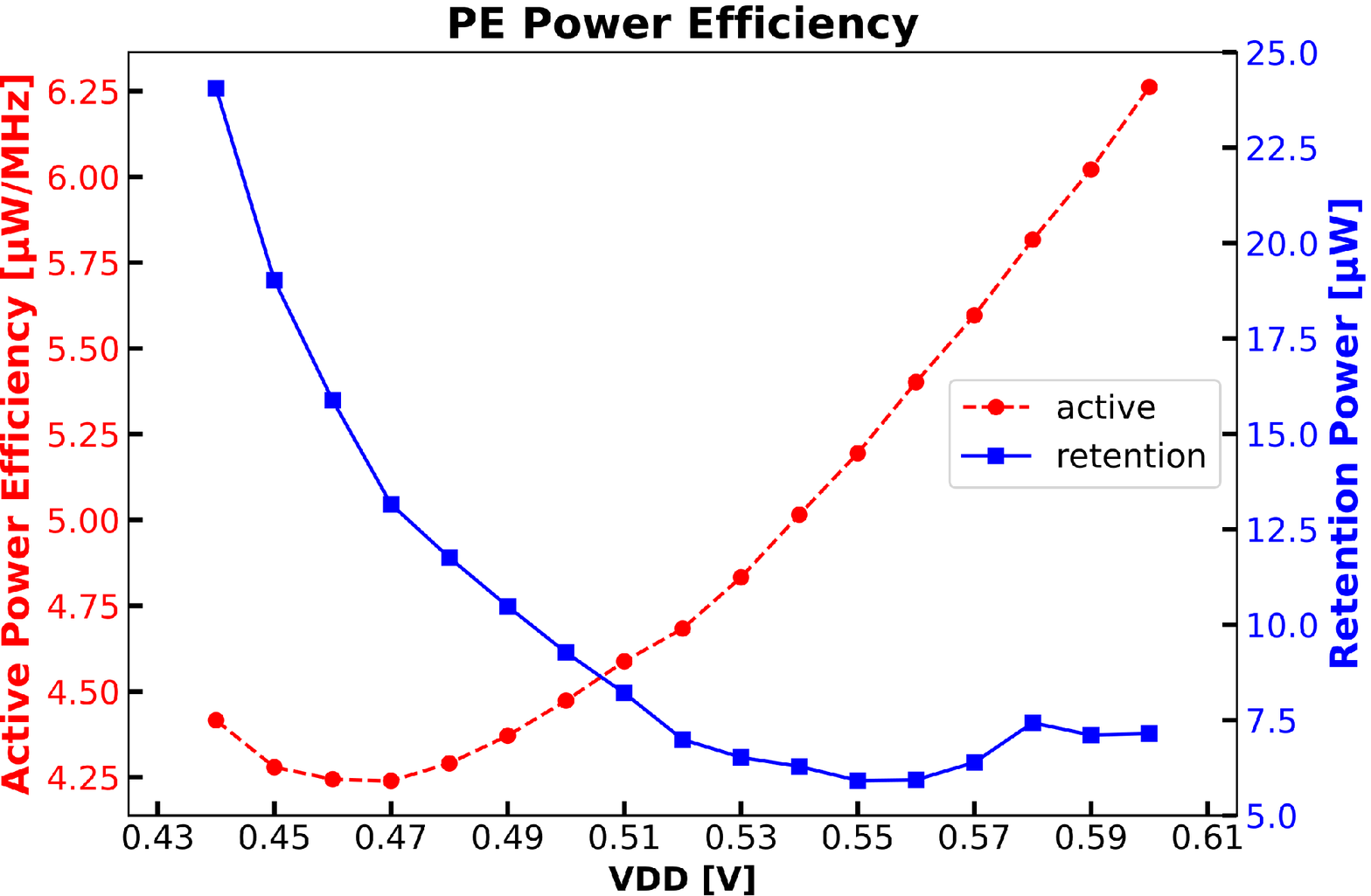}}
\caption{PE power efficiency running CoreMark @ 25~MHz and PE sleep power during retention sleep mode.}
\label{fig:pe_coremark_ret_plt}
\end{figure}

\begin{figure}[t]
\centerline{\includegraphics[width=85mm,keepaspectratio]{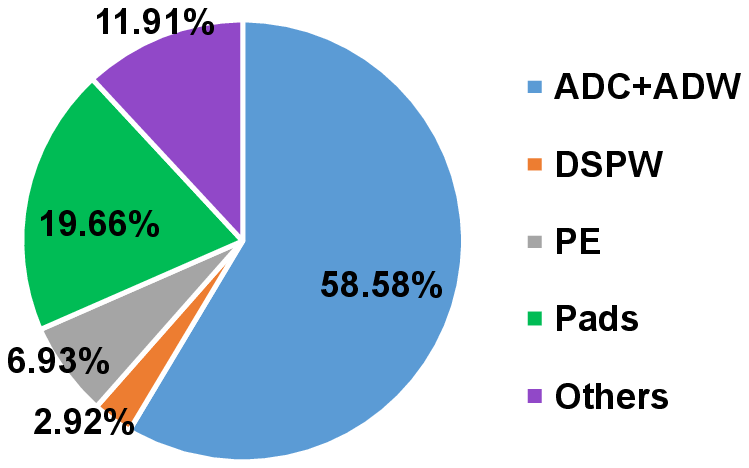}}
\caption{Chip area break-down. }
\label{fig:chip_area_bd}
\end{figure}

Figure~\ref{fig:pe_coremark_ret_plt} shows the PE power efficiency running CoreMark at various VDD voltages. We achieve 5.19~µW/MHz at a nominal VDD of 0.55~V. By reducing VDD to 0.47~V the efficiency is 4.24~\textmu W/MHz. As shown in  Figure~\ref{fig:pe_coremark_ret_plt} the retention sleep power consumption has its minimum of 5.91~\textmu W at nominal VDD. A reduction of VDD causes the ABB regulation to adapt the back bias voltage to meet speed requirements resulting in higher cell leakage. As previously introduced, the PE offers significant on-chip programmability, facilitating tasks such as training for CCE and both training and inference for spike sorting. 
The on-chip training of CCE consumes 16.94~\textmu J. Meanwhile, the on-chip training of a software-based spike sorting process requires 2.4~nJ/sample. Similarly, the inference process incurs an energy consumption of 1.43~\textmu J/spike. With the utilization of the MAC accelerator, energy consumption is enhanced by 23.8\% to 1.09~\textmu J/spike.

The chip area break-down is shown in Figure~\ref{fig:chip_area_bd}. It depicts that the AFE plus ADW dominates the area. Conversely, the area occupied by the DSPW is minimal in comparison. Consequently, the proposed PSoC imposes a modest area overhead when juxtaposed with traditional neural implants. This observation underscores the scalability and adaptability of the proposed on-chip processing engines.

\section{Discussion}\label{sec:discussion}
\subsection{Comparison}\label{subsec:compare}
A comparison with other state-of-the-art neural signal acquisition and processing chips is presented in Table \ref{tab:comparison}. From the comparison, it is evident that the proposed PSoC exhibits a greater integration of on-chip neural processing approaches. Additionally, for each individual approach, our PSoC offers comparable or superior results.

\begin{table*}[btp]
    \caption{Comparison with state-of-the-art designs}\label{tab:comparison}
	\centering
	\begin{threeparttable}
    \resizebox{\linewidth}{!}{
	\begin{tabular}{|c|>{\columncolor[gray]{0.9}}c|c|c|c|c|c|}
	\hline 
		\multirow{2}{*}{Works} & This & A-SSCC~2022 & JSSC~2022 & TBioCAS~2022 & ISSCC~2023 & JSSC~2022 \\
		& work & \citep{schueffny2022} & \citep{uran2022} & \citep{zeinolabedin2022} & \citep{chen2023} & \citep{shin2022} \\
	\hline 
		Tech. (nm) & 22 & 22 & 65 & 22 & 22 & 65\\
	\hline 
		\multirow{2}{*}{Volt. (V)} & AFE: 0.4, & AFE: 0.4, & \multirow{2}{*}{1.0} & AFE: 0.8, & AFE: -, & AFE: 1.2, \\
        &  Dig.: 0.55 & Dig.: 0.55 & & Dig.: 0.5-0.8 &  Dig.: 0.59 & Dig.: - \\
	\hline 
		Number of AFE ch. & 68 & 64 & 16 & 16 & No & 256 \\
	\hline
		Compression & AP \& LFP & AP & AP \& LFP & No & No\tnote{\dag} & No \\
	\hline
		\multirow{2}{*}{Spike detection} & 2-stage (Adap. & 2-stage (Adap. & \multirow{2}{*}{Hard thresholding} & \multirow{2}{*}{NEO} & \multirow{2}{*}{NEO} & \multirow{2}{*}{No} \\
        & Thr. \& NEO) & Thr. \& NEO) & & & & \\
  	\hline 
		\multirow{2}{*}{Feature extraction} & PCA / AF / & \multirow{2}{*}{No} & \multirow{2}{*}{CHT} & \multirow{2}{*}{AF} & \multirow{2}{*}{Peak-FSDE} & \multirow{2}{*}{HT / BPF} \\
        & SW-based & & & & & \\
	\hline
		Spike sorting & SW-based & No & No & Adap. SC & Geo-OSort & No \\
	\hline
		ATE area (mm$^2$) & 0.0031 & 0.0058 & No & No & No & No \\
	\hline
	    \multirow{3}{*}{CE-SSR} & AP: 63\% (LL) / & AP: 62.5\% (LL) / & AP / LFP: & \multirow{3}{*}{No} & \multirow{3}{*}{No\tnote{\dag}} & \multirow{3}{*}{No} \\
        & 91\% (NLL), & 91\% (NLL), & 80\% & & & \\
        & LFP: 64\% (LL) & LFP: No & (Lossy) & & & \\
	\hline 
		CE-Power/Ch. & AP: 0.87 - 4.39, & AP: 16.47 - 17.84, & AP \& LFP: & \multirow{2}{*}{No} & \multirow{2}{*}{No\tnote{\dag}} & \multirow{2}{*}{No} \\
        (\textmu W) & LFP: 0.32 & LFP: No & 1.83 & & & \\
    \hline 
		CE-Area/Ch. & AP: 0.0026 / 0.0064, & AP: 0.147, & AP \& LFP: & \multirow{2}{*}{No} & \multirow{2}{*}{No\tnote{\dag}} & \multirow{2}{*}{No} \\
        (mm$^2$) & LFP: 0.00043 & LFP: No & 0.0076 & & & \\
    \hline 
		\makecell{PE-Energy (CE-\\Training, \textmu J)} & 16.94 & No & No & No & No & No \\
    \hline 
		\makecell{PE-Energy (SS-\\Training, \textmu J)} & 15.36 & No & No & 28.46 & No & No \\
    \hline
        SS-Accuracy & 91.48\% / 94.12\% \tnote{\ddag} & No & 92\% / 97.8\% \tnote{\P} & 94.12\% & 89.5\% & 95.6\% \\
    \hline
        \multirow{2}{*}{SS-Datasets} & Quiroga & \multirow{2}{*}{No} & CHB-MIT & \multirow{2}{*}{Quiroga} & \multirow{2}{*}{Quiroga} & \multirow{2}{*}{CHB-MIT} \\
        & \citep{quiroga2004} & & \citep{shoeb2010} & & & \\
    \hline
        AFE-Power (\textmu W) & 0.41 & 0.40 & 0.65 & 1.52 & No & 1.51 \\
    \hline
	\end{tabular}
}
\begin{tablenotes}
	\footnotesize
	\item[\dag] The spike sorting is categorized as a form of compression in this paper. However, it is only the index of spiking neurons and is not reconstructible.
    \item[\ddag] Two distinct feature extraction algorithms are employed here. 
    \item[\P] The accuracy here was performed off-chip. 
\end{tablenotes}
\end{threeparttable}
\end{table*}

In the proposed PSoC, 68 recording frontends are integrated, surpassing those in \cite{schueffny2023, zeinolabedin2022, uran2022}, thereby enabling the recording of a larger number of neural signal channels. Furthermore, the power consumption of 0.41~\textmu W, encompassing both the analog frontend and digital wrapper, also demonstrates advantages compared to almost all state-of-the-art alternatives. 

The on-chip digital signal processing wrapper in this work demonstrates remarkable versatility and efficiency. For AP compression, our PSoC offers two modes: The lossless mode and the near-lossless mode. In the lossless mode, neural signals can be perfectly reconstructed, whereas in the near-lossless mode, the position and waveform of all spikes can be reconstructed without loss. In comparison to \cite{uran2022}, our approach achieves a higher SSR of 91\% in the near-lossless mode, while strictly preserving spike waveforms. Additionally, compared to \cite{schueffny2023}, although the SSR is similar, our compression engine exhibits significant advantages in terms of area and power consumption per channel, making it highly applicable. 
For LFP compression, despite having a lower SSR than \cite{uran2022}, our PSoC holds three distinctive features: 1) significantly smaller area, 2) substantially lower power consumption, and 3) preservation of signal reconstruction integrity with no loss. 
Our spike detector operates in two stages, offering greater stability compared to the hard thresholding method employed in \cite{uran2022}, as well as the single NEO stage documented in \cite{zeinolabedin2022, chen2023}. When compared to the 2-stage SD used in \cite{schueffny2023}, our adaptive threshold estimator occupies only half the area. 
Furthermore, the spike sorting capability of our PSoC achieves a satisfactory accuracy of 91.48\% over the Quiroga datasets in \cite{quiroga2004}, which is comparable to the accuracies achieved by other methods using the same datasets. 
Additionally, our PSoC includes an efficient FIR filter and spike raster, expanding its range of potential applications.

Ultimately, the integrated ultra-low power MAC-assisted PE offers programmability and versatility for a multitude of applications, including on-chip training for CCE, as well as on-chip training and inference for spike sorting. Compared to \cite{zeinolabedin2022}, our PSoC achieves lower energy consumption for tasks such as spike sorting.
z

\subsection{Conclusion}\label{subsec:conclu}
In this paper, we introduce a highly-integrated neural signal processing PSoC fabricated using 22nm FDSOI technology, occupying a compact chip size of 9~mm$^2$. Leveraging advanced ABB technology, our design achieves remarkable ultra-low power consumption.
The architecture of our PSoC encompasses 68-channel AFEs, a digital processing wrapper, and a MAC-assisted PE. The digital processing wrapper features a robust command-based central control unit and functional units, facilitating comprehensive on-chip processing of neural signals with exceptional energy efficiency. This includes tasks such as lossless/near-lossless compression of AP, lossless compression of LFP, filtering of raw neural signals for LFP extraction, generation of spike raster plots, and adaptive estimation of spike detection thresholds to enhance the performance of the spike detection module. 
Notably, the AFE consumes a mere 0.25~\textmu W/Ch, while the ADW consumes only 0.16~\textmu W, housing digital filters, spike detectors, and FIFO buffers. 
The proposed PSoC achieves an SSR of approximately 91\% for AP in the near-lossless mode and an SSR of 64\% for LFP losslessly, consuming only 0.87 (GC) - 4.39 (AC)~\textmu W/Ch and 0.32~\textmu W/Ch, respectively. The 16-channel FIR filter exhibits an average power consumption of 25.76~\textmu W. 
Furthermore, the on-chip MAC-assisted PE offers programmability and versatility for various applications, including on-chip training for ICE and CCE, as well as on-chip training and inference for spike sorting. 
The on-chip training for the CCE module consumes an average of 16.94~\textmu J. The on-chip training of a software-based spike sorting requires 2.4~nJ/sample. Correspondingly, the inference process consumes 1.43~\textmu J/spike. By employing the bio-specific MAC accelerator, the energy consumption is improved by 23.8\% to 1.09~\textmu J/spike, achieving an average accuracy rate of 91.48\% or 94.12\% based on the utilized features. The PE consumes a power of 5.19~\textmu W/MHz at a supply voltage of 0.55~V. By reducing the supply voltage to 0.47~V, the efficiency is 4.24~\textmu W/MHz. The retention sleep mode has a power consumption of approximately 5.91~\textmu W.
Compared to conventional neural implants, the PSoC proposed in this work significantly reduces data rate and transmission power consumption, while offering diverse on-chip processing capabilities, thereby enabling the potential increase in the number of recording channels for a new class of cortical active and intelligent implants.

\section*{Conflict of Interest Statement}

The authors declare that the research was conducted in the absence of any commercial or financial relationships that could be construed as a potential conflict of interest.

\section*{Author Contributions}

LG, SMAZ, and RG conceived the study. LG contributed to implementing, verifying, and measuring most digital components and performed spike sorting. AW implemented the FIR filter and MAC unit and conducted in-vitro validation. SMAZ implemented the top-level of the PSoC and digital components in the analog frontend. FMS contributed to the analog frontend. MS implemented the PE block, assisted in setting up power measurements, and contributed to STM firmware. QM and ZW contributed to implementing compression engines. SS, AD, DW, GE, and SH did back-end work for chip implementation including constraining and sign-off verification. MB was responsible for PCB design. JP assisted in algorithm analysis. RG led the project, contributed to in-vitro validation and STM firmware. CM supervised the findings of this work. All authors discussed the results and contributed to the final manuscript.

\section*{Funding}
This work was supported by European Union's Horizon 2020 Programme (``SYNCH", GA n. 824162) and the Federal Ministry of Education and Research of Germany in the programme of ``Souverän. Digital. Vernetzt.". Joint project 6G-life, ID: 16KISK001K.

\section*{Acknowledgments}
The Article Processing Charge (APC) were funded by the joint publication funds of the TU Dresden, including Carl Gustav Carus Faculty of Medicine, and the SLUB Dresden as well as the Open Access Publication Funding of the DFG.

\section*{Data Availability Statement}
The Quiraga datasets used in this study for AP compression and spike sorting can be found \href{https://figshare.le.ac.uk/articles/dataset/Simulated_dataset/11897595}{here}. 
The Buzsáki datasets used in this study for LFP compression can be found \href{https://buzsakilab.nyumc.org/datasets/}{here}.

\end{document}